\documentclass{aa}  
\usepackage{xcolor}
\usepackage{graphicx}
\usepackage[varg]{txfonts}
\usepackage{isotope}
\usepackage{siunitx}
\usepackage{ulem}
\usepackage{bm}
\usepackage{natbib}
\usepackage{url}
\usepackage{xspace}
\usepackage{placeins}
\usepackage{capt-of}
\usepackage{booktabs}
\usepackage{subcaption}  

\bibpunct{(}{)}{;}{a}{}{,}

\usepackage[colorlinks=true,citecolor=blue,linkcolor=magenta,urlcolor=blue]{hyperref}
\usepackage[switch,columnwise]{lineno}

\usepackage{float}
\newcommand{\gstar}{J1903-0023}
\newcommand{\gstarfull}{Gaia\,DR3\,4265540383431508736}

\newcommand{\gaia}{\textit{Gaia}}
\newcommand{\wise}{J0725-2351}
\newcommand{\wisefull}{WISE J072543.88-235119.7}
\newcommand{\feh}{[Fe/H]}
\newcommand{\alp}{[$\alpha/\mathrm{Fe}$]}
\newcommand{\msun}{$M_{\odot}$}
\newcommand{\rsun}{$R_{\odot}$}

\newcommand{\teff}{$T_\mathrm{eff}$}
\newcommand{\logg}{$\log g$}
\newcommand{\kms}{km\,s$^{-1}$}
\newcommand{\uh}[1]{\textcolor{magenta}{#1}}
\newcommand{\synthe}{\textsc{Synthe}}
\newcommand{\atlas}{\textsc{Atlas}{\footnotesize12}}

\begin{document}

   \title{Ancient, eclipsing, tidally-locked: A blue lurker progenitor in the population of extreme-velocity star candidates}

   \author{A. Bhat
          \inst{1}
          \and
          M. Dorsch
          \inst{1}
          \and
          S. Geier
          \inst{1}
          \and
          U. Heber
          \inst{2}
          \and
          H. Dawson
          \inst{1}
          \and
          F.Mattig
          \inst{1}
          \and
          D. Benitez-Palacios
           \inst{3}
          \and
          Pablo Fernandez-Schlosser
          \inst{3}
          }

   \institute{Institut für Physik und Astronomie, Universität Potsdam, Haus 28, Karl-Liebknecht-Str.\ 24/25, 14476 Potsdam, Germany\\
              \email{aakashbhat7@gmail.com}
         \and
             Dr.\ Karl\ Remeis-Observatory \& ECAP, Astronomical Institute, Friedrich-Alexander University Erlangen-Nuremberg, Sternwartstr.~7, 96049 Bamberg, Germany
             \and
             Instituto de Física y Astronomía, Universidad de Valparaíso, Gran Bretaña 1111, Playa Ancha, Valparaíso 2360102, Chile
             }


 
  \abstract
{Many extreme velocity candidate stars have been found based on \textit{Gaia} astrometry, but need spectroscopic confirmation.
We select late-type hypervelocity star (HVS) candidates from the \textit{Gaia} DR3 catalog with a $1\sigma$ lower limit of the tangential velocity of 800 km\,s$^{-1}$. \gstar, one of the brightest targets, stands out as high priority candidate for follow-up spectroscopy using the X-shooter instrument at ESO-VLT. We determine its atmospheric parameters and abundances utilizing synthetic spectral grids and a global $\chi^2-$minimization procedure, and its stellar parameters with the help of evolutionary tracks and the spectral energy distribution. The star shows variability in its light curve and follow-up spectroscopy confirms that the star is radial-velocity variable. The spectroscopic distance of \gstar\ is lower than that based on the parallax, indicating that the star is not a hypervelocity binary star but bound to the Galaxy.
The star turned out to be of spectral type F, very similar to the extreme-velocity star \wise, which we analyse in the same way as the target.
Apparently, both stars are very metal poor and old halo main-sequence (sdF) stars with masses slightly below the halo turn-off mass, and share the low metallicity ([Fe/H]=-2.3,-2.6) and strong alpha enhancement ([$\alpha$/Fe]$\sim0.44$).
While \wise\ is non-rotating ($v\sin\,i<3$\,km\,s$^{-1}$), \gstar\ is a fast rotator ($v\sin i=42.3\pm2.0$ km\,s$^{-1}$).  The \gaia\, and ZTF light curves show an eclipse at a 1.179 day period, similar to the rotation period of \gstar. We therefore conclude that \gstar\ is a high-velocity tidally-synchronised binary most likely with a metal-poor M dwarf companion.}

   \maketitle
%
\section{Introduction}
Old, extreme-velocity halo stars that are tracers of the history of Galactic formation are now being discovered by combining large spectroscopic surveys with \textit{Gaia} astrometry \citep{Li2023}. Many such stars were originally candidate hypervelocity stars (HVS) unbound to the Galaxy, whose ejection mechanisms are still unclear, but they turned out to be bound when precise astrometry from \gaia\ became available \citep{2024AJ....167..225W,Boubert2018}. The lack of radial velocity measurements calls for follow-up spectroscopy. For example, \citet{Hawk2018}, \citet{Reggi22}, and \citet{nelson2024} have studied a total of 36 late-type HVS candidates allowing them to find elemental abundances and study the formation scenarios of such stars. These scenarios include, but are not exclusive to, accretion from satellite galaxies, in-situ formation through dynamic heating and ejections from the disc. Stars formed through these mechanisms may have similar kinematic properties. Precise studies of spectra to characterize the metallicity and individual abundances of the stars provide valuable information to constrain the nature of the stars.

In this letter, we report the identification of the candidate HVS \gstarfull\ (\gstar\, from here on), as a tidally synchronised binary in the halo. The system will undergo mass-transfer soon and can be characterized as a pre-blue lurker system. The term blue lurker has been coined for stars that are the lower-mass counterparts of blue stragglers. They show higher rotation rates than main-sequence stars of the same mass and appear to be younger than their predicted age estimates in clusters \citep{2019ApJ...881...47L}.
The overall binary fraction as well as the tight binary fraction of metal poor stars is a highly debated topic. \citet{2002AJ....124.1144L} found that the observed fraction of binaries was roughly similar in the disc and the halo. Recent studies show that the close binary fraction is likely anti-correlated with metallicity \citep{Moe2019}. \citet{Bashi2024} also find a significant fraction of binaries in accreted and in-situ halo stars.

The spectroscopic analysis (section 3), light curve (section 5), and spectral energy distribution (SED) (section 6) of the star suggest that it is a metal-poor F-type star, most likely with an M-dwarf companion. We also analysed one of the X-shooter spectra of the extreme-velocity sdF-type star \wisefull\, (\wise\, from here) which was reported by \citet{Scholz2015}, as a twin comparison star to validate our results.

\section{Target selection, spectroscopic and photometric observations}\label{sect:obs}

We searched for fast stars using the same query to the \gaia\, DR3 catalog \citep{2023A&A...674A...1G} as \citet{kareemfast}, who used this method to identify several hypervelocity white dwarfs. To search for main sequence stars, the colour criterion was extended to include $G_\mathrm{BP}$\,$-$\,$G_\mathrm{RP}$>$0.7$\,mag. Among all objects, \gstar\ stands out as the only star with a tangential velocity high enough to indicate HVS kinematics.  The star has a parallax of $0.40 \pm0.06$ mas, making it one of the closest known HVS candidate. The high proper motion of the star translates into a Galactocentric tangential velocity of $\sim800$\,km\,s$^{-1}$. The astrometric properties of \gstar\ were investigated by \citet{scholz2024}, who applied relevant quality criteria to 72 stars whose Galactic transversal velocities exceed 700\,km\,s$^{-1}$. \gstar\ passed all checks except for the nearest-neighbour criterion because it is located in a crowded field in the Galactic plane, where parallax uncertainty might be up to 4 times larger\citep{scholz2024}. \gstar\ is in the top 10 highest priority extreme velocity stars and among the brightest (G=16.5 mag). 

We combined photometric magnitudes to construct the SED and fitted it with synthetic models as described in \citet{Uli2018} to get an estimate of the angular diameter, reddening, and stellar temperature. The SED had no IR excess, and hinted at a highly reddened F-type star (see Fig.~\ref{fig:gaiaSED}). Combining the \gaia\, parallax with the angular diameter, we calculated a radius of $1.5 \pm 0.2$ $R_\odot$ corresponding to a mass of $\sim 1.7$ \msun, for a log $g\sim4.5$.
A precise measurement of the atmospheric parameters, metallicity, and radial velocity was required to probe further. The star was later reported to be variable in \citet{Gaia3}, although with only a few data points.

We obtained optical spectra at the ESO VLT with the X-shooter instrument at high signal-to-noise ($\mathrm{S/N}>60$) in the blue and VIS arms of X-shooter ($\lambda = 3000 - 9000\, \text{\AA}$ at a resolving power $R = 5500 \,\text{to}\, 11500$). Due to the light curve variability in the Zwicky Transient Facility (ZTF) light curve, two low S/N, low resolution spectra were obtained to check for radial velocity (RV) variability from the Nordic Optical Telescope (NOT/ALFOSC)\footnote{https://www.not.iac.es/instruments/alfosc/} and Southern Astrophysical Research (SOAR/Goodman)\footnote{https://noirlab.edu/public/programs/ctio/soar-telescope/goodman/} spectrographs  covering the optical range from approximately 3600 \,\AA\, to 5200 \AA. Using a 1" slit, each setup achieves a spectral resolution of 2.2\,\AA, and 2.6\,\AA, respectively, corresponding to RV precisions between 10 and 15 \kms. 
For \wise\, we reanalyzed an archival X-Shooter spectrum ($R$ $= 6655$ in the blue arm, and $R$ $=8935$ in the VIS arm), of high S/N ($165$ in the blue, $178$ in the VIS).

\section{Spectroscopic analysis}

We used the model grids from the \atlas\ and \synthe\ codes \citep{Kurucz1996}, which assume local thermodynamic equilibrium (LTE) to compute synthetic spectra. The
extended model grids cover a metallicity range [Fe/H] of -3 to 0.5 and an alpha enhancement [$\alpha$/Fe] of 0 to 0.45. After removing cosmics and data reduction artifacts, the entire spectrum was
fit using a global $\chi^2$- minimization routine, similar to that described in \citet{Irrg2014}. We simultaneously fit the log g, Teff, [Fe/H], [$\alpha$/Fe], the projected rotational velocity $v\,\sin{i}$ and the radial velocity of the star. Poorly modeled spectral regions
were removed from the fit.

\begin{figure}
\centering
\includegraphics[width=0.5\textwidth]{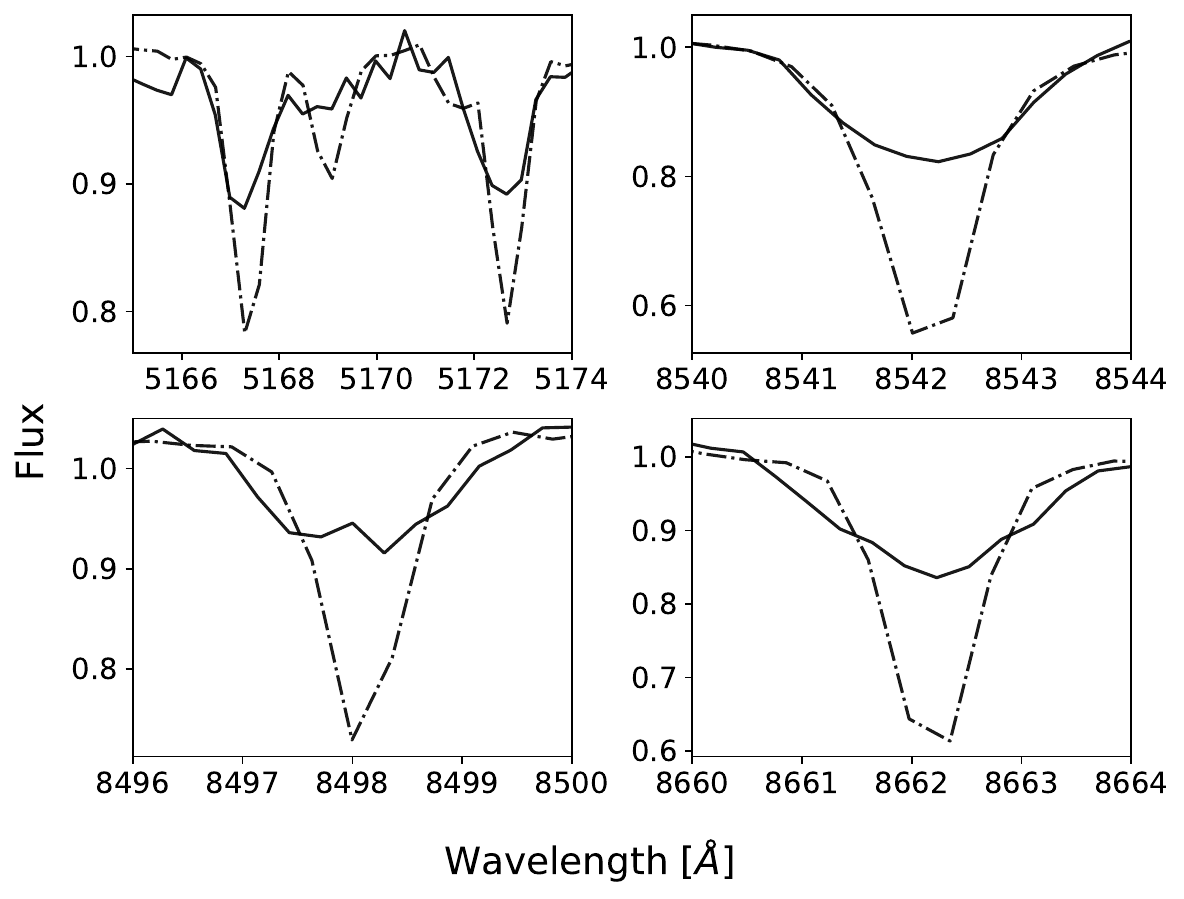}
\caption{The Mg\,I and Ca\,II triplets in the normalized and RV-corrected spectra of \gstar\ (solid) and \wise\ (dash-dotted). 
 The rotational velocity broadening can be clearly seen for \gstar. 
The resolving power is 6655 for the blue arm and 8935 for the VIS arm for \wise\ and the spectrum of  \gstar\ was convolved to match it.  
 }
\label{fig:metals}
\end{figure}

The spectral fits are shown in Figs.\ \ref{fig:2} and \ref {fig:scholz1}.  The stars have very similar $T_{\mathrm{eff}}$, log g, metallicities, and alpha enhancements. \wise\ is known to be very metal poor \citep{Scholz2015} and we derive [Fe/H]=$-2.33\pm 0.10$ with strong alpha enhancement ([$\alpha$/Fe]=$0.44\pm 0.05$). \gstar\ is even more metal poor at \feh\ $\approx -2.63$ with the same large alpha enhancement as \wise\, which implies that both stars are likely old and best classified as population II subdwarfs (sdF). 
The most intriguing difference, however, is the projected rotational velocity. While \wise\ is non-rotating as expected for an ancient halo star, \gstar\ is rotating unusually fast at $v\sin i=42\pm2$ km/s. To demonstrate the difference, 
Figure \ref{fig:metals} compares the Mg\,{\sc i} and Ca\,{\sc ii} triplets in the normalised spectra of \wise\ and \gstar; while the lines are sharp for \wise\ all spectral lines of \gstar\ are significantly broadened by rotation.

The 
very low metallicity is at odds with our previous assumption of the SED-based classification, because its radius must be much smaller and, therefore, the star is closer than indicated by the \gaia\, parallax, which 
leads us to believe that the \gaia\, parallax is underestimated and cannot be used to derive a reliable stellar radius and mass. Further analysis of spectra from NOT/ALFOSC and SOAR/Goodman for \gstar\, was done by fitting the same model with the atmospheric parameters fixed to the Xshooter fit. An RV shift of $-80 $ \kms\, was found for both spectra, consistent with the \gaia\, classification as an eclipsing binary (see appendix \ref{A.spectra}).

\begin{table}[h!]
\renewcommand{\arraystretch}{1.1}
\centering
\caption{Best fit parameters for X-shooter spectrum fitted using a $\chi^2$ minimization method. Statistical uncertainties are provided.}
\begin{tabular}{c c c}
\hline
\hline
Parameter& \gstar & \wise\\
\hline
$T_{\mathrm{eff}}$ [K]& $6416 \pm 130$ & $6320 \pm 120$  \\
log $g$ & $4.32 \pm 0.15$ &$4.46 \pm 0.10$ \\
 \text{[Fe/H]} & $-2.63 \pm 0.10$ &$-2.33\pm 0.10$\\
$v\sin i$ [km\,s$^{-1}$] & $42.3 \pm 2.0$ & $0\pm3$ \\
\text{[$\alpha$/Fe]}  & $0.447 \pm 0.050 $& $0.44 \pm 0.05$\\
$v_{r}$ [km\,s$^{-1}$] & $-120 \pm 2$& $240 \pm 2$ \\
\hline
\end{tabular}
\label{tab:params1}
\end{table}

\section{Stellar masses and ages {from evolutionary tracks}}\label{sect:evol}

 We derive estimates of masses and ages for the stars 
based on MIST evolutionary tracks \citep{MIST0,MIST1,mesa1,mesa2,mesa3} for metallicities \feh\, $-2$, $-2.5$, and $-3.0$. Utilizing a Bayesian MCMC approach, we put priors on the mass (less than $1.6$ M$_\odot$, that is twice the halo turn-off mass), age (less than $14$ Gyr), and metallicity (Gaussian centered at our observed metallicity) and we interpolate to sample near the log $g$ and T$_{\rm eff}$ we derive from the spectra. We use the SciPy package \citep{2020SciPy-NMeth} along with the Emcee package \citep{emcee} for this purpose. This allowed us to get the posterior distributions of log $g$, $T_{\text{eff}}$, age, 
and  metallicity. For the case of \wise\, we find a median mass 
of $0.74^{+0.03}_{-0.03} M_\odot$ and an age of $9.77^{+2.77}_{-4.41} \text{Gyr}$.
The stellar mass is lower than  \citet{Scholz2015}, due to a higher log $g$ derived from the new model grid. The mass derived from tracks shown in Fig. \ref{fig:interp_scholz} is in agreement with the spectro-photometric mass using \textit{Gaia} parallax.
 
We repeat the procedure for our new sdF star to see if a single star solution exists for it. We find no conclusive parameters for this star. Fig \ref{fig:interp_gaia} shows that the derived parameters are inconsistent with evolutionary tracks for ages smaller than 14 Gyr at $1\sigma$ confidence level. Models with rotation $\sim 0.4\,V_{c}$, corresponding to $\sim 140$ km s$^{-1}$ for a $0.8$ M$_\odot$ star do not change the results. Within $2\sigma$ uncertainty the star may be young ($<1$ Gyr) with a mass of $0.75$\,$M_\odot$, or old ($\sim13$ Gyr) with a mass of $0.72$\,\msun. The younger age is highly unlikely for a halo star and is possibly due to the large uncertainty on the \logg. The final adopted parameters of the two stars are given in Tab. \ref{tab:params2}. 

\section{Light Curve Analysis}\label{sect:lc}
\begin{figure}
\centering
\includegraphics[width=0.4\textwidth]{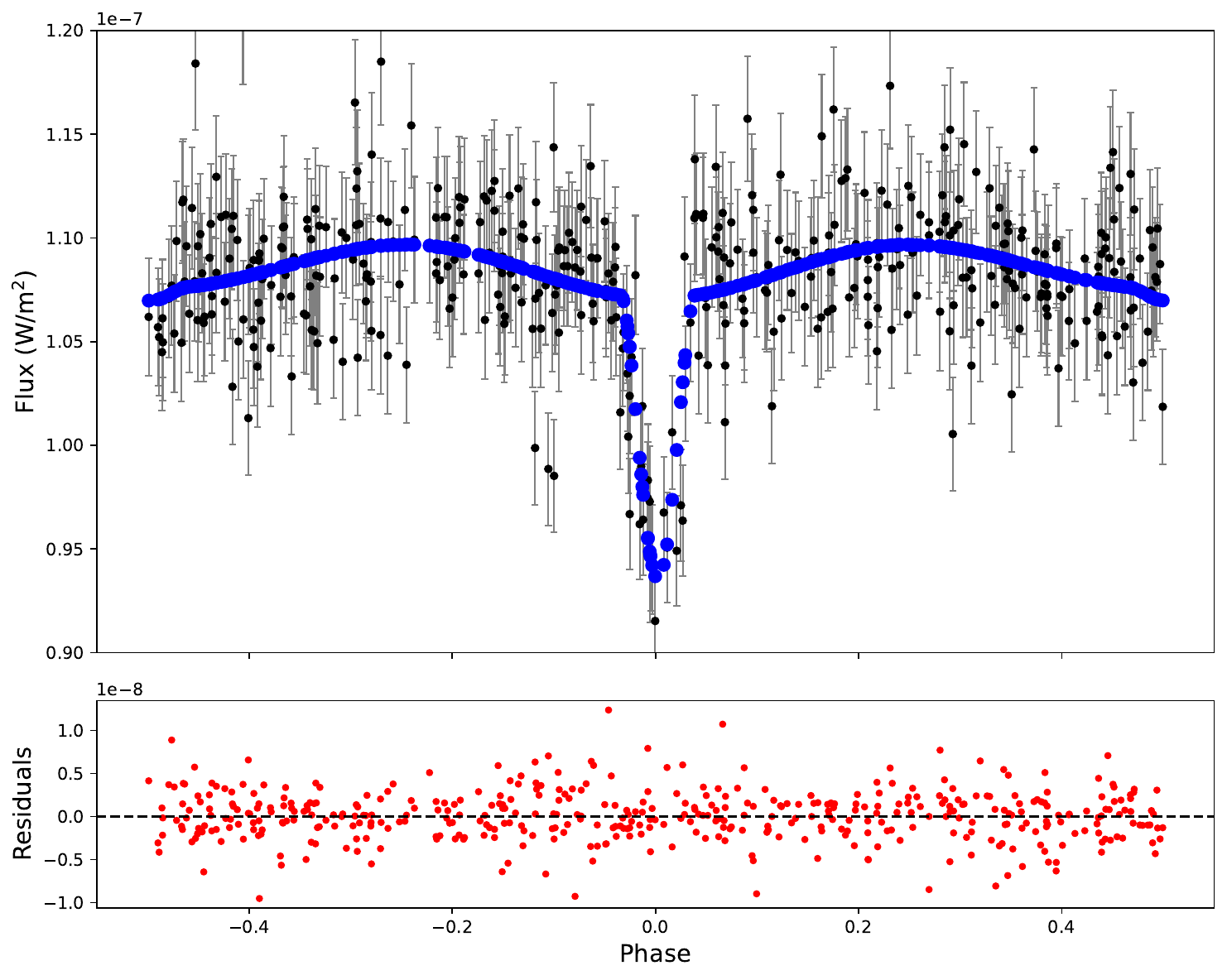}
\caption{ZTF LC phase-folded with a period of 1.179 days. The blue line is the phoebe model.
 }
\label{fig:ZTF_lc}
\end{figure}

\gstar\, was reported by \gaia\, to be an eclipsing binary star with a period of 1.179 days \citep{2022yCat.1358....0G}, although the light curve is quite noisy and has many period peaks. However, the star also has a ZTF light curve \citep{ztf1,ztf2}, which allowed us to constrain its binary parameters. Since no flux from the secondary is detected in the SED and no secondary eclipses can be discerned in the light curve, the companion must have low luminosity. Using the inverse modelling as decribed by \citet{2020ApJS..250...34C}, we were able to model the g-band light curve with a primary with the spectro-photometric \teff\, and radius (see Tab. \ref{tab:params2}), and a secondary component with $3940$ K and radius of $0.46$ \rsun\, in PHOEBE \citep{phoebe1,phoebe2,phoebe3} and conclude that this is the most likely architecture of the binary. Spectroscopic and photometric follow-up is needed to solve the system. The phase-folded observed light curve and the model are shown in Fig. \ref{fig:ZTF_lc}. The parameters were then sampled using the EMCEE package \citep{2013PASP..125..306F} with a gaussian prior on the secondary radius centered at 0.5 \rsun\, (see Tab.~\ref{tab:lc_parameters}), and Appendix \ref{A.lcmodel}.

\begin{table}[h!]
\centering
\caption{Median light curve Parameters with $1\sigma$ uncertanties}
\begin{tabular}{ll}
\hline
\textbf{Parameter} & \textbf{Value} \\
\hline
Period ($P$) & 1.179 $\pm$ 0.005 days \\
\teff (Secondary) & $3560^{+590}_{-830}$ K\\
Inclination & $82.0^{+3.6}_{-3.2}$ deg\\
Mass Ratio (q) & $0.47^{0.19}_{-0.16}$\\
R$_{\text{equiv}}$ (Secondary) & $0.36^{+0.16}_{-0.05}$\rsun\\
\hline
\end{tabular}
\label{tab:lc_parameters}
\end{table}

\section{Radii, luminosities, and distances 
}\label{sect:sed}

We repeated  the SED analysis of \gstar\ and \wise\
described in Sect. \ref{sect:obs} but used the spectroscopic values of $T_{\mathrm{eff}}$, [Fe/H] and log $g$, 
with the same model grids as in the spectral analysis to obtain their angular diameters and line-of-sight reddening\uh{.} 
The resulting best-fit models are compared to the observed SEDs in Figs.\ \ref{fig:gaiaSED} and \ref{fig:scholzSED}, and the corresponding values are listed in Table \ref{tab:params2}. The models match the observations very well. There is no hint of the presence of a cool companion that might show up in the IR. An M dwarf with a temperature of $3560$ K would contribute $<1.7\%$ to the luminosity and is therefore hard to detect in the SED.

\begin{table}[h!]
\small
\renewcommand{\arraystretch}{1.1}
\centering
\caption{The combined spectro-photometric parameters for the two stars.}
\begin{tabular}{c c c}
\hline
\hline
Parameter& \gstar & \wise\\
\hline
Angular diameter $\log \Theta$ & $-10.512 \pm 0.006$& $-9.827 \pm 0.002$ \\
Color excess $E(44-55)$& $0.547 \pm 0.012$ & $0.041 \pm 0.004$  \\
Evolutionary mass [\msun]& $0.75\pm 0.03$ &$0.74\pm0.03$ \\
Radius ($R=\sqrt{GM/g}$) [$R_\odot$]  & $0.99 \pm 0.09 $& $0.84 \pm 0.05$\\
Distance d$_{\rm spectro}$ [kpc] & $1.46 \pm 0.1$ &$0.250\pm 0.03$\\
$v_{\text{tan}}$ [km\,s$^{-1}$] & $585 \pm 40$& $318 \pm 38$ \\
Luminosity [$L_\odot$]& $1.40\pm 0.39$ & $0.99\pm0.19$\\
\hline
\end{tabular}
\label{tab:params2}
\end{table}

\gstar\, is highly reddened with a colour excess  $E$(44-55) of $0.547\pm0.012$ mag because of its location in the Galactic disc ($b$=-2.8$^\circ$). \wise\ lies in the disc ($b$=-3.6$^\circ$), but its reddening is low because it is much closer. The high reddening is visible in the spectrum of \gstar\, in the form of diffuse interstellar band features at $4430, 6177, 6284, 6614, 8620\,\text{\AA}$ (see Fig~\ref{fig:2}).
The stellar masses have been derived by comparing the atmospheric parameters to predictions from evolutionary tracks (see Sect.~\ref{sect:evol}).
Hence, we can derive the stellar radii from their masses and gravities and their distances from angular diameter and radius (see Table \ref{tab:params2}). Finally, the luminosities are derived from the effective temperatures and radii.

While the spectroscopic distance for \wise\ is consistent with the \gaia\ based parallax, for \gstar\ the former (d$_{\rm spectro}=1.45 \pm 0.1$ kpc) is significantly smaller than the latter (d$_\varpi=2.22^{+0.37}_{-0.29}$ kpc). This adds doubt as to the reliability of the \gaia\ parallax \citep[see][]{scholz2024} and results in a more moderate heliocentric tangential velocity of  $585 \pm 40$ km/s$^{-1}$ than the previous estimate (see also Sect. \ref{sect:kinematic}). This discrepancy is further evident in the difference in parallax from DR2 (0.19 mas) to DR3 (0.4 mas). 

\section{Kinematics}
\label{sect:kinematic}

We calculate trajectories from spectroscopic distances, 
and radial velocities as described in \citet{2018Andreas,andreas}. We use a revised version of the model introduced by \citet{1991RMxAA..22..255A}, called Model 1, for the Galactic potential and numerically integrate the equations of motion for the stars using a $4$th-order Runge-Kutta solver. The local standard of rest (LSR) velocities (U,V,W)$_\odot$ are $(11.1\pm1, 12.24\pm2,7.25\pm0.5)$ \kms\, taken from \citet{2010MNRAS.403.1829S}.

For \gstar\, it is difficult to measure the radial velocity of the system, since the three spectra for a binary with a 1.178 day period capture a small phase space. Using the ephemeris and period obtained from the light-curve, we get a $\gamma = -130 \pm 40$ \kms\, with a slightly inflated uncertainty to capture the uncertainty in the ephemeris and the RVs. Since the star's tangential velocity component is the dominant one the RV doesn't affect the kinematics significantly. Running multiple models with different radial velocities, the star can only be unbound if the systemic velocity is $<-600$ \kms\, or $>250$ \kms. These velocities are highly unlikely since the maximum peak to peak velocity can only be close to $\sim180$ \kms\, making the star bound. 

\begin{figure}
\centering
\includegraphics[width=0.45\textwidth]{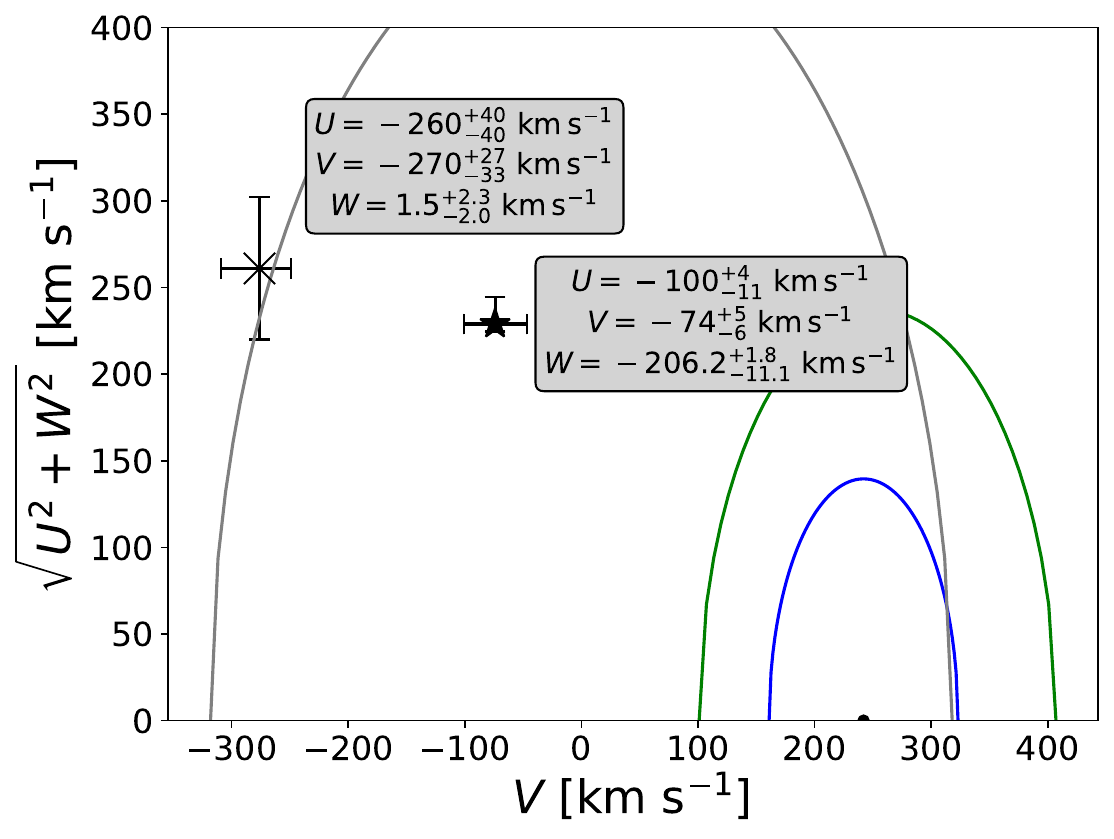}
\caption{The Toomre diagram for \gstar\ (cross), and \wise\ 
 (star). 
The $3\sigma$ contours of the thin disc (blue), the thick disc (green), and the halo (grey) are constructed from the $U$, $V$ and $W$ distributions of \citet{angiano2020}. $U$ is the velocity component towards the Galactic center, $V$ in the direction of Galactic rotation, and $W$ perpendicular to the Galactic plane.
}
\label{fig:toomre_lz_E}
\end{figure}

Both stars are on retrograde orbits, strengthening the conclusion that they belong to the halo population. 
The proper motion of \wise\ ($268.11\pm0.02$ mas yr$^{-1}$) is in agreement with that measured by \citet[][$266.7\pm1.9$ mas yr$^{-1}$]{Scholz2015} derived from ground-based proper motions, but much more precise than the latter. 
However, because the parallax-based distance is smaller than the  previously adopted one, its Galactic restframe velocity decreases from $v_{\text{grf}} = 460$ km s$^{-1}$ to $240$ km s$^{-1}$. 
\gstar\, is bound to the Galaxy but is an extreme velocity object with the velocity vector lying at the edge of the $3\sigma$ contours of the halo in the Toomre diagram as shown in Fig. \ref{fig:toomre_lz_E}. 

\begin{figure}[htbp]
\centering
    \begin{subfigure}{0.47\linewidth}
        \centering
        \includegraphics[width=\linewidth]{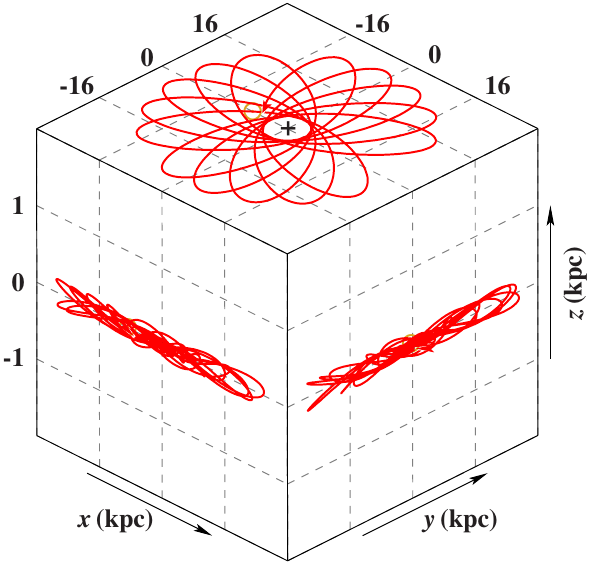}
    \end{subfigure}
    \hfill
    \begin{subfigure}{0.47\linewidth}
        \centering
        \includegraphics[width=\linewidth]{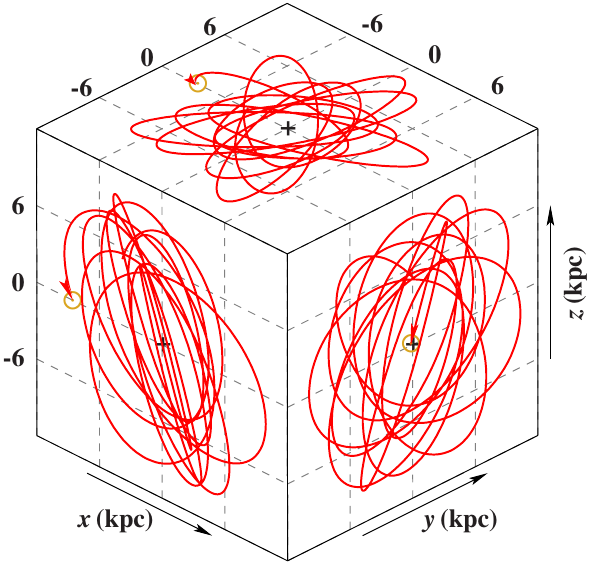}
    \end{subfigure}
\caption{Galactic orbits for \gstar\ (left panel) and \wise\ (right panel), in a Galactic Cartesian coordinate system with the z-axis pointing to the Galactic North Pole, calculated back in time for $15$ Gyr. The Galactic centre is marked with a plus sign, the location of the Sun at $x=-8.33\pm0.35$ kpc is shown as an orange circle.
\gstar\, has a peculiar trajectory with no significant motion in the Z-direction, 
while that of \wise\ is characteristic for a halo star.  }
\label{fig:kin_bound}
\end{figure}

The Galactic trajectories of the two stars are very different (see Fig. \ref{fig:kin_bound}).  The eccentric orbit ($e=0.38^{+0.05}_{-0.01}$) of \wise\ is typical for halo stars as it extends to large distances in all coordinates, that is, it is a halo star caught when passing through the Galactic plane as suggested by \citet{Scholz2015}. The orbit of \gstar\ is even more eccentric ($e=0.72^{+0.06}_{-0.06}$) and extends far out in the plane to $R_{\text{max}}\sim 19$ kpc, but stays close to it over the entire integration time. 
 Potential Galactic origins of the stars are discussed in the appendix, which might be responsible for their different kinematics. We conjecture that \gstar\ might be related to the Phlegethon stream or the Sequoia merger.

\section{Evolutionary status}

Our analysis of \wise\ confirms the conclusion of \citet{Scholz2015} that it is an ancient halo main sequence star.
The masses of both stars are slightly lower than the halo turn-off mass. Such stars in globular clusters do not rotate, as shown by \citet{2003Lucatello}. \gstar\ shares the fast rotation with the lower-mass counterparts of blue stragglers, called blue lurkers
\citep{2019ApJ...881...47L,2020JApA...41...45S,2023MNRAS.523L..58D}.

However blue lurkers are formed either through the mergers of two low-mass stars, stellar collisions in dense environments, mass transfer through wind Roche-lobe overflow \citep{sun2024}, or through a stable mass transfer episode \citep{2019ApJ...881...47L}. Blue lurkers formed by mass transfer need to gain mass from their companions, which are required to have evolved past the MS phase to increase their radii and fill their Roche lobes. Since the companion in our system is suspected to be an unevolved M dwarf, such a scenario is highly unlikely. 

The short period and old age of the system implies that tidal effects lead to a circularisation of the orbit and the  synchronisation of the components' rotation periods and the orbital period \citep{tidal1,tidal2,tidal3}. The rotation period of the sdF, using inclination from the light curve, the spectrophotometric radius, and the fitted $v\,\sin{i}$, comes out to $1.190\pm0.12$ days. This is the same period as observed for the orbital period of the system, suggesting that the system is synchronized. 

Due to the short period, the Roche-lobe radius of the system is $\sim2$ \rsun\ using the Eggleton approximation \citep{1983ApJ...268..368E}. Extending the MIST tracks further, the star will fill its Roche-lobe within the next $1-3$ Gyr and start the first phase of mass-transfer. Depending on how conservative the mass transfer will be, the M dwarf companion will gain mass and turn into a blue lurker or even blue straggler. 

\section{Summary and conclusion}
We reported the spectral, photometric, light-curve, and kinematic analysis of the hyper-velocity candidate \gstar, having one of the highest tangential velocities in \textit{Gaia} DR3, and a reanalysis of the high-velocity sdF star \wise\ for reference. Using X-shooter$@$ESO-VLT spectra we derive atmospheric parameters, abundances, and projected rotation and radial velocities. Using the ZTF light-curve and multiple spectra we confirm the \gaia\, variability, which lines up well with an M dwarf companion in a 1.179 day orbit.
 Comparing spectroscopic \teff\ and \logg\ measurements with MIST evolutionary tracks, we derive very similar masses of $0.74\pm0.03$ M$_\odot$ and $0.75\pm0.03$ M$_\odot$ for \wise\ and \gstar, respectively. Despite their location in the Galactic plane,
both objects are low metallicity ($-2.3$ to $-2.6$) and alpha enhanced ([$\alpha$/Fe]=$0.44$). With their retrograde orbits, they both belong to the Galactic halo, where the close binary fraction of accreted stars has been found to be higher than in-situ ones \citep{2024MNRAS.535..949B}. 

The spectroscopic distance of \wise\ is consistent with its \gaia\ parallax, but not of \gstar\, which is significantly closer than indicated by its parallax. The \gaia\, DR3 parallax is likely underestimated because \gstar\ is in a close binary and because it lies in a crowded field \citep[see][for a discussion]{scholz2024}. Accordingly, its tangential velocity is smaller, and the star is bound to the Galaxy. The Galactic trajectories of the stars differ enormously, with \wise\ showing an orbit typical for halo stars, while that of \gstar\ is unusual because it stays close to the Galactic plane and is very eccentric. We conjecture that its origin lies within a Galactic stream. 

The most intriguing difference between the stars is in their rotation.
While \wise\ is non-rotating ($v\sin i<3$\,km\,s$^{-1}$), \gstar\ is a fast rotator ($v\sin i=42.3\,\pm2.0$ km\,s$^{-1}$) with a period that aligns well with its orbital period. The latter is very unusual for an isolated ancient sdF star, but the rotation period matches the orbital period and we conclude that tidal interaction with a close M dwarf companion has spun up the star.  With a period of around 1.179 days, \gstar\, is one of the tightest known main-sequence binary in the halo \citep[see][for periods of such binaries where circularization takes place around 20 days, in comparison to which \gstar\, is a on a much shorter orbit] {2002AJ....124.1144L} and provides a new avenue to study such systems in the field.

\begin{acknowledgements}
We thank the anonymous referee for their comments which helped us better understand this system. We also thank Dr. Veronika Schaffenroth for their help with the light curves.
A.B. was supported by the Deutsche Forschungsgemeinschaft (DFG) through grant GE2506/18-1. M.D. was supported by the Deutsches Zentrum für Luft- und Raumfahrt (DLR) through grant 50-OR-2304. D.B. acknowledges financial support from ANID-Subdirección de Capital Humano/Doctorado Nacional/2025-21250735. Based on observations collected at the European Southern Observatory under ESO programmes 112.26Z3.001 and 093.D-0127. 
Fitting procedures were done on the clusters of University of Potsdam, and Dr.\ Karl-Remeis Sternwarte, Bamberg. 
This work has made use of data from the European Space Agency (ESA) mission
{\it Gaia} (\url{https://www.cosmos.esa.int/gaia}), processed by the {\it Gaia}
Data Processing and Analysis Consortium (DPAC,
\url{https://www.cosmos.esa.int/web/gaia/dpac/consortium}). Funding for the DPAC
has been provided by national institutions, in particular the institutions
participating in the {\it Gaia} Multilateral Agreement. This research has made use of the VizieR catalogue access tool, CDS,
Strasbourg, France \citep{vizier2} and the SIMBAD database,
operated at CDS, Strasbourg, France \citep{simbad}. The data presented here were obtained [in part] with ALFOSC, which is provided by the Instituto de Astrofisica de Andalucia (IAA) under a joint agreement with the University of Copenhagen and NOT. Based in part on observations obtained at the Southern Astrophysical Research (SOAR) telescope, which is a joint project of the Minist\'{e}rio da Ci\^{e}ncia, Tecnologia e Inova\c{c}\~{o}es (MCTI/LNA) do Brasil, the US National Science Foundation’s NOIRLab, the University of North Carolina at Chapel Hill (UNC), and Michigan State University (MSU).
\end{acknowledgements}
\bibliographystyle{aa}
\footnotesize
\bibliography{references}

\clearpage
\appendix
\section{Spectra}
\label{A.spectra}
\begin{figure*}
\centering
\includegraphics[width=0.9\textwidth]{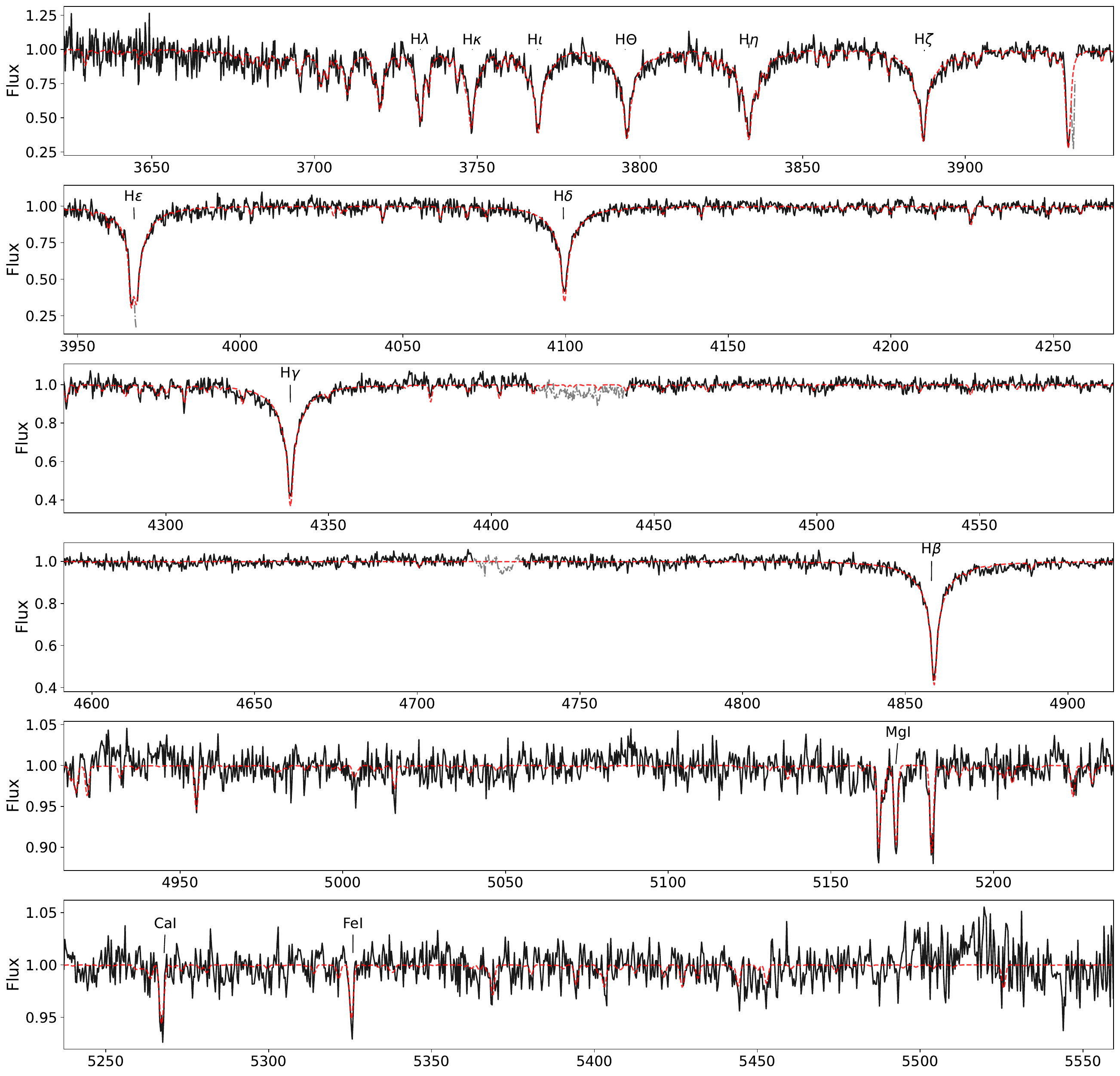}
\includegraphics[width=0.9\textwidth]{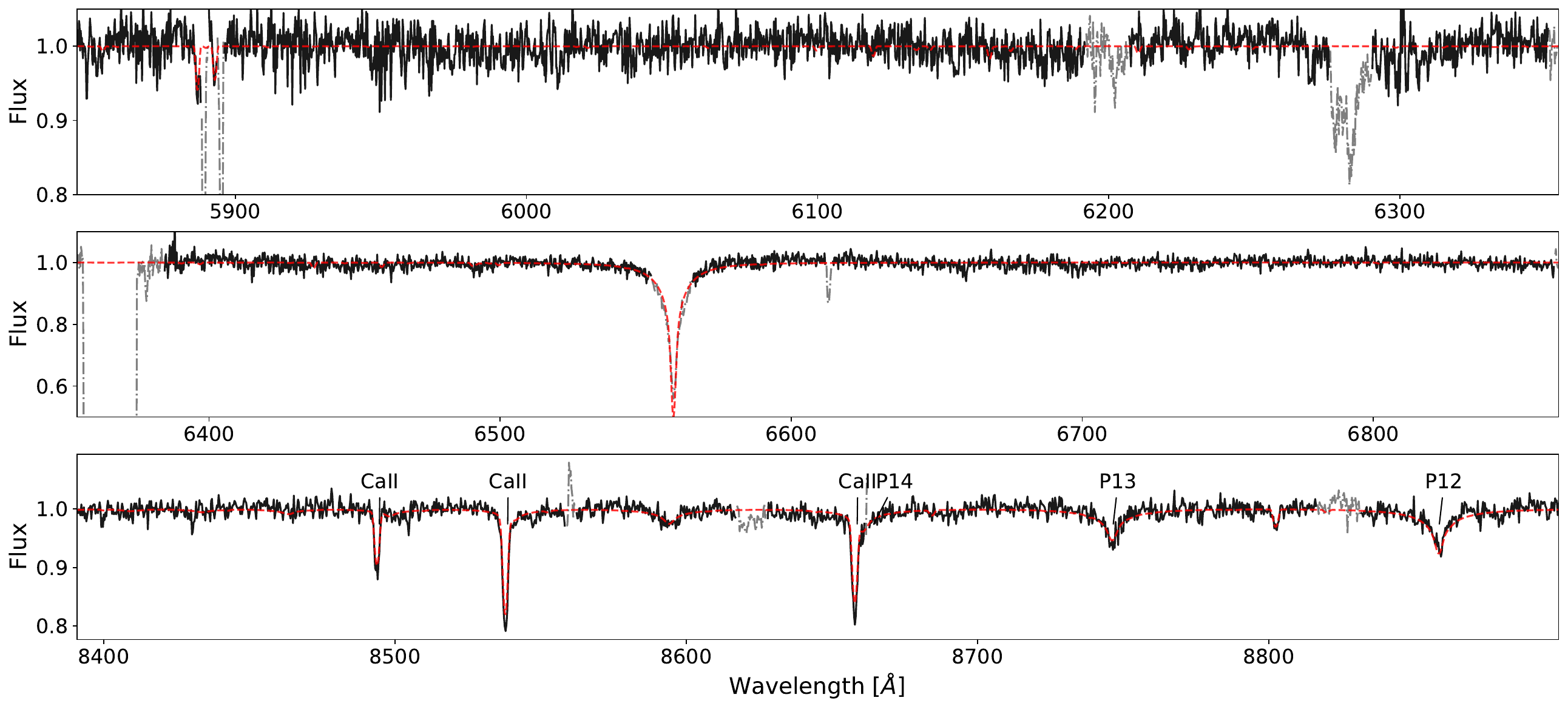}
\caption{X-shooter spectrum from the UVB and VIS arms with best fit model. The normalized flux is shown with solid black lines and the best fit model is shown in dashed red lines. Parts of the spectra which were not considered in the fit are marked with the dashed black line and include diffuse interstellar bands due to the high reddening. Panels have different flux ranges. The strongest lines of hydrogen, magnesium, calcium, and iron are also marked. Paschen lines are marked in the infrared, and some of them blend with Ca\,II lines. }
\label{fig:2}
\end{figure*}

\begin{figure}
\centering
\includegraphics[width=0.45\textwidth]{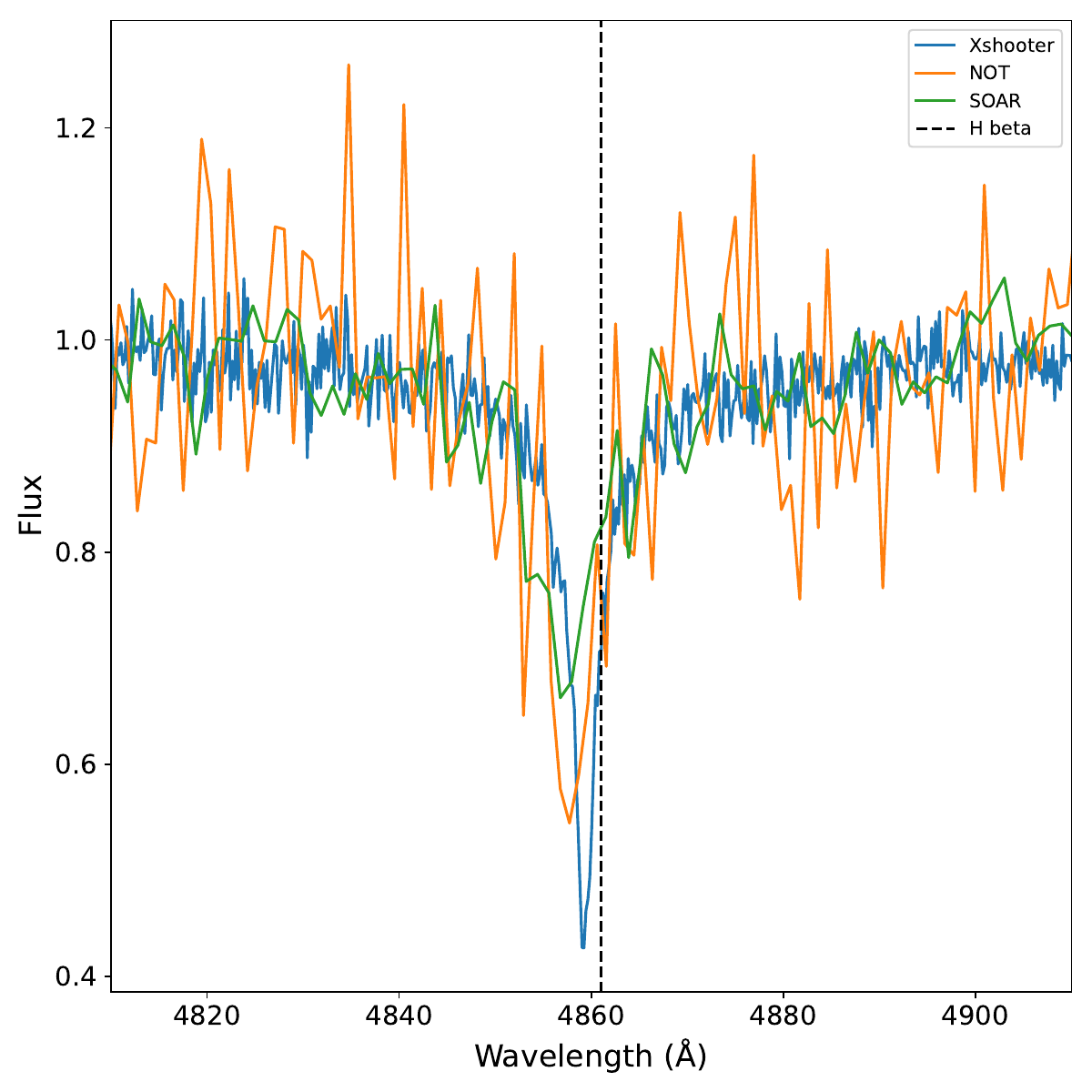}
\includegraphics[width=0.45\textwidth]{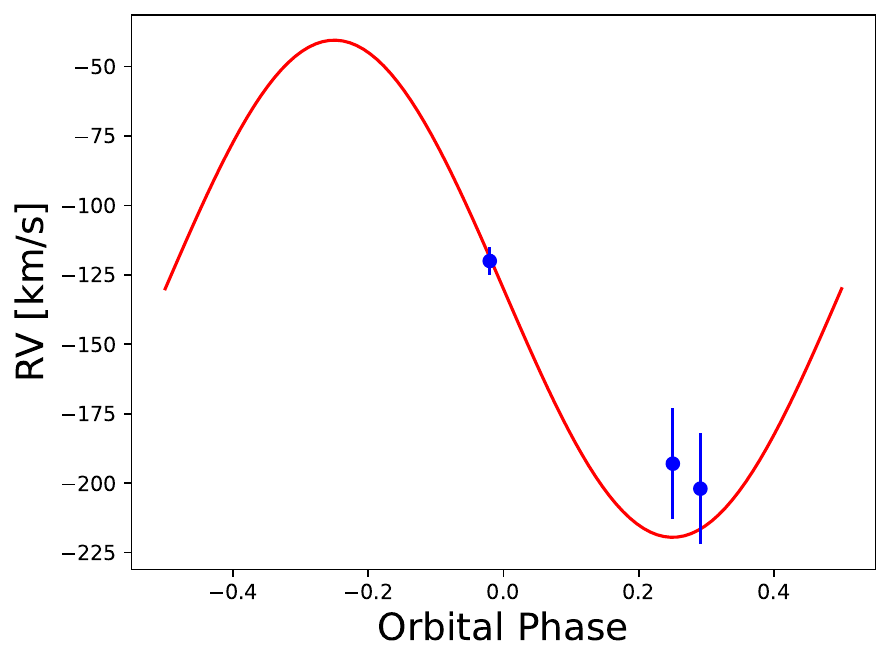}
\caption{[Top]: The normalized H beta (dashed black line) is shown for comparison from Xshooter (blue), NOT (orange), and SOAR (green). The spectral resolutions are different for all three. [Bottom]: The phase-folded RV curve with $\gamma=-130$ \kms.}
\label{fig:comp}
\end{figure}

The X-Shooter spectrum of \gstar\, is shown in Fig. \ref{fig:2}. We model this as a single star spectrum with no visible contamination through a secondary component. Table \ref{Tab:compspec} shows the different RVs for the system. Fig. \ref{fig:comp} shows the difference in RV. 
The spectra obtained with NOT/ALFOSC and SOAR/Goodman were reduced using an open-source pipeline available on GitHub\footnote{\url{https://github.com/Fabmat1/SOAR_data_reducer}}. The reduction process included standard bias subtraction and flat-field correction. Wavelength calibration was performed using arc lamp exposures taken immediately prior to each science observation.

\begin{table}[h!]
\centering
\caption{Radial velocity measurements from different instruments. Uncertainties are systematic.}
\begin{tabular}{lcc}
\hline
\textbf{Instrument} & \textbf{MJD} & \textbf{RV (km/s)} \\
\hline
Xshooter & 60401.37822708035& $-120 \pm 3$ \\
NOT/ALFOSC  &60787.2301286805 & $-196 \pm 20$ \\
SOAR/GOODMAN & 60820.29014157407 &$-203 \pm 20$ \\
\hline
\end{tabular}
\label{Tab:compspec}
\end{table}

\begin{figure*}
\centering
\includegraphics[width=0.9\textwidth]{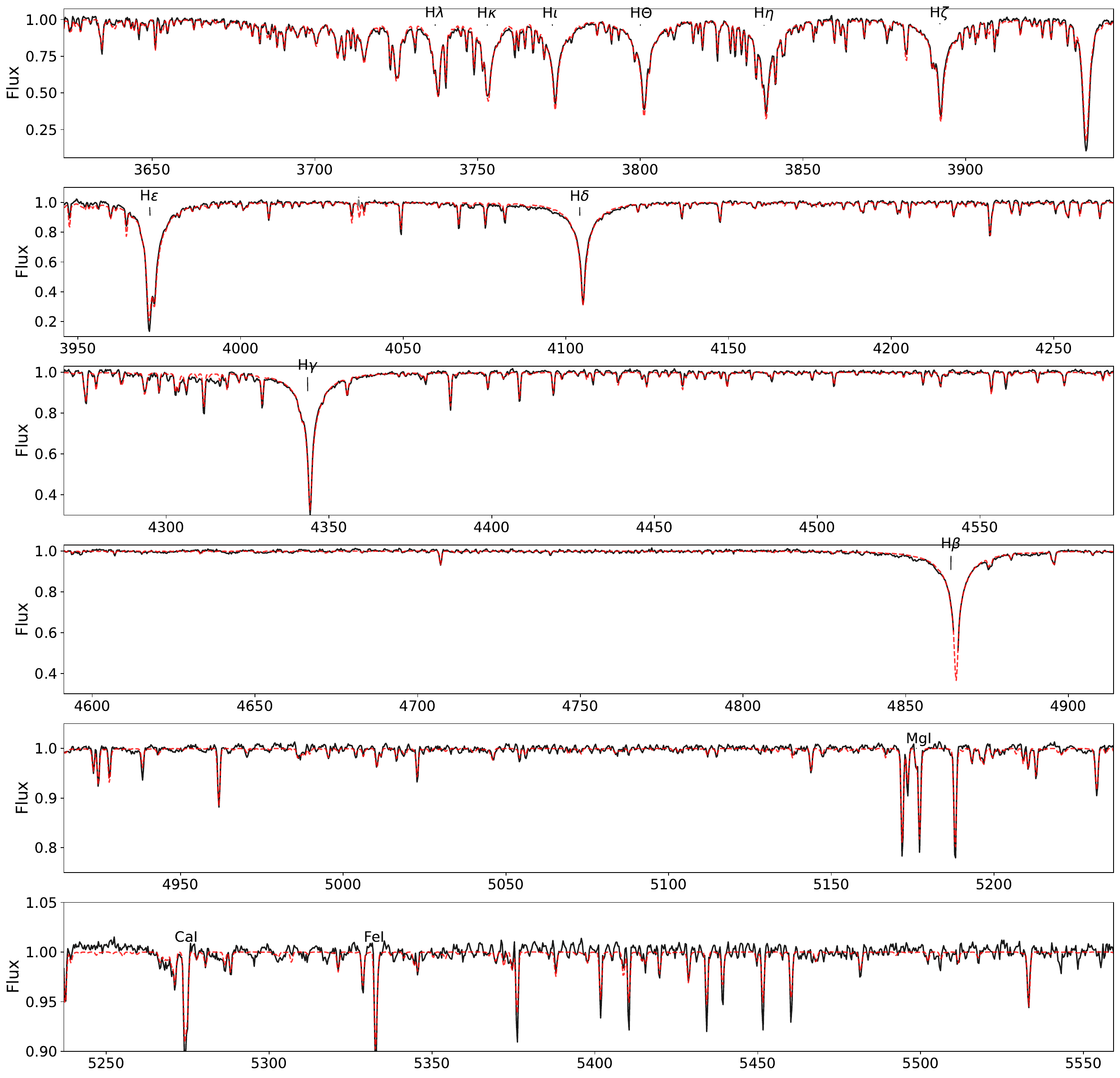}
\includegraphics[width=0.9\textwidth]{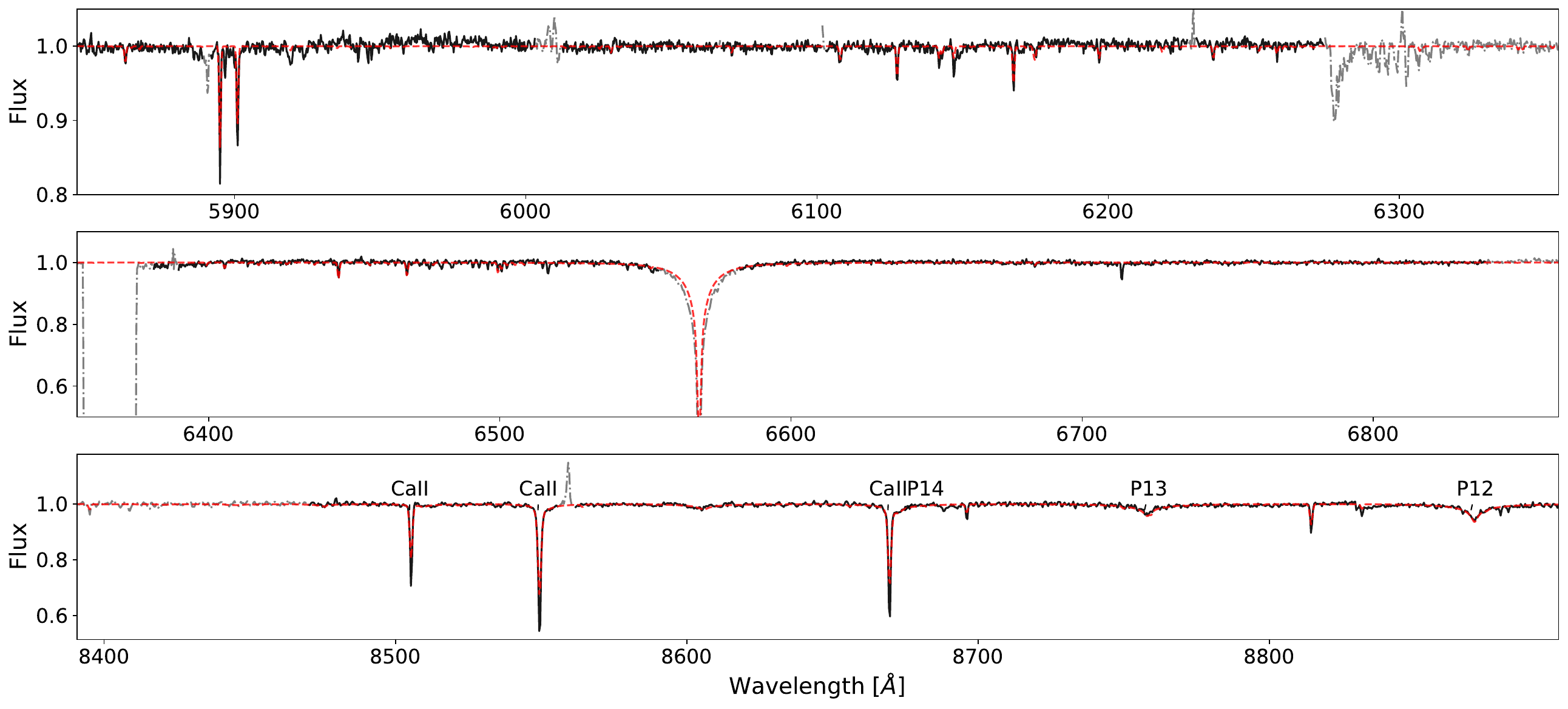}
\caption{X-shooter spectrum of \wise\,from the UVB and VIS arms with best fit model fit. Refer to Fig. \ref{fig:2} for explanations.}
\label{fig:scholz1}
\end{figure*}

\section{Tracks and future evolution}
\label{B.Tracks}

\begin{figure}
\centering
\includegraphics[width=0.45\textwidth]{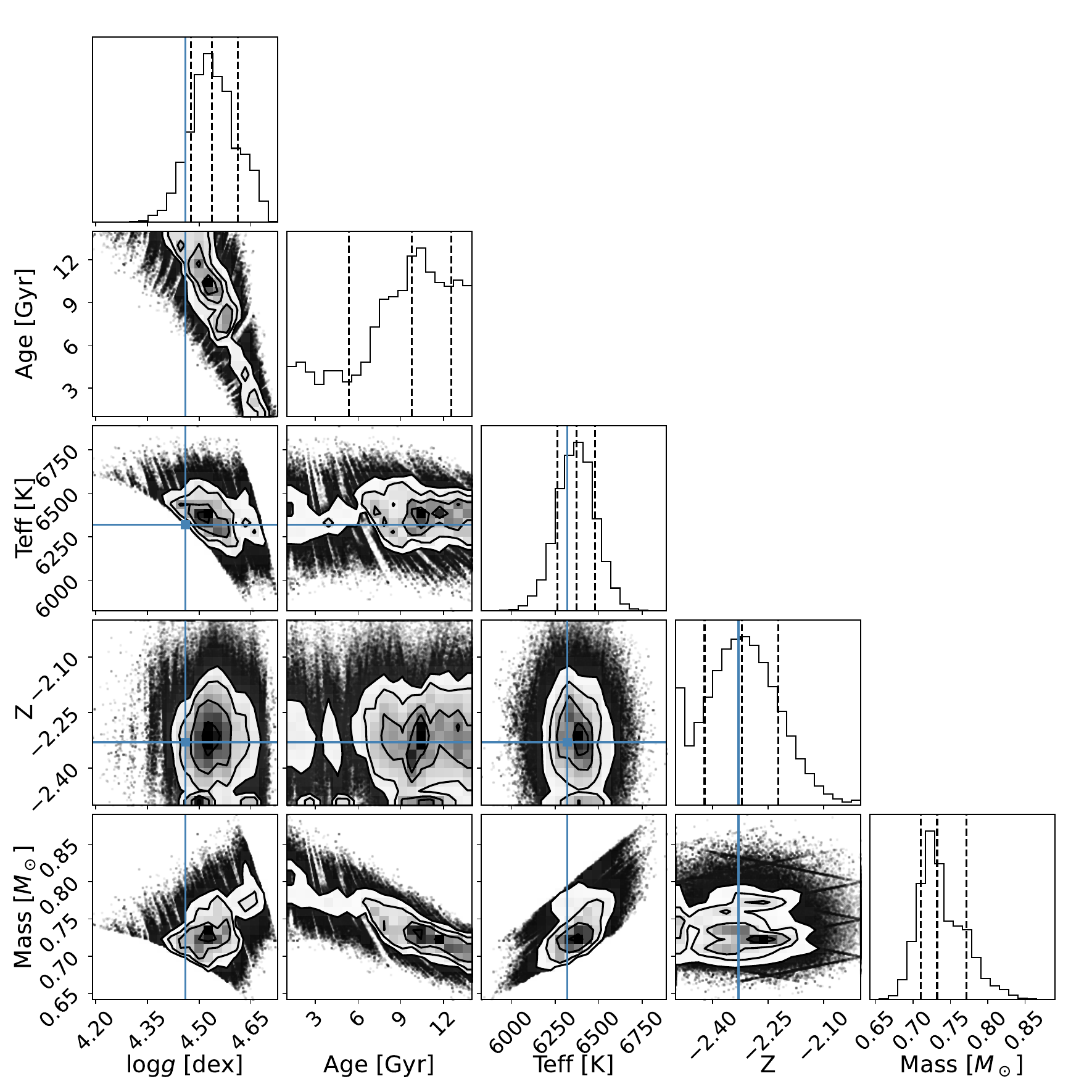}
\includegraphics[width=0.45\textwidth]{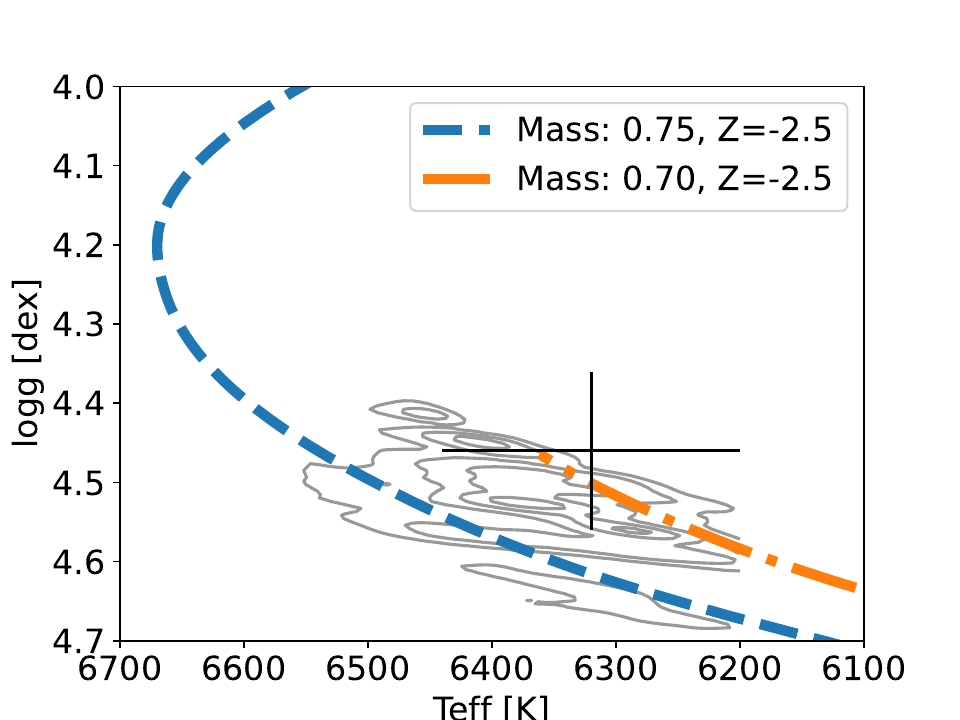}
\caption{Corner plots and evolutionary tracks for stellar parameters of \wise. \textbf{upper panel}: Posterior distribution of the interpolated values for log $g$, Age, \teff, metallicity, and mass. The blue lines in the corner plot show the observed values for the star. \textbf{lower panel}: Tracks are shown for comparison for $0.75$ \msun\, and $0.70$ \msun\, at a metallicity of $-2.5$. Grey contours show the sampled distribution. }
\label{fig:interp_scholz}
\end{figure}

\begin{figure}
\centering
\includegraphics[width=0.45\textwidth]{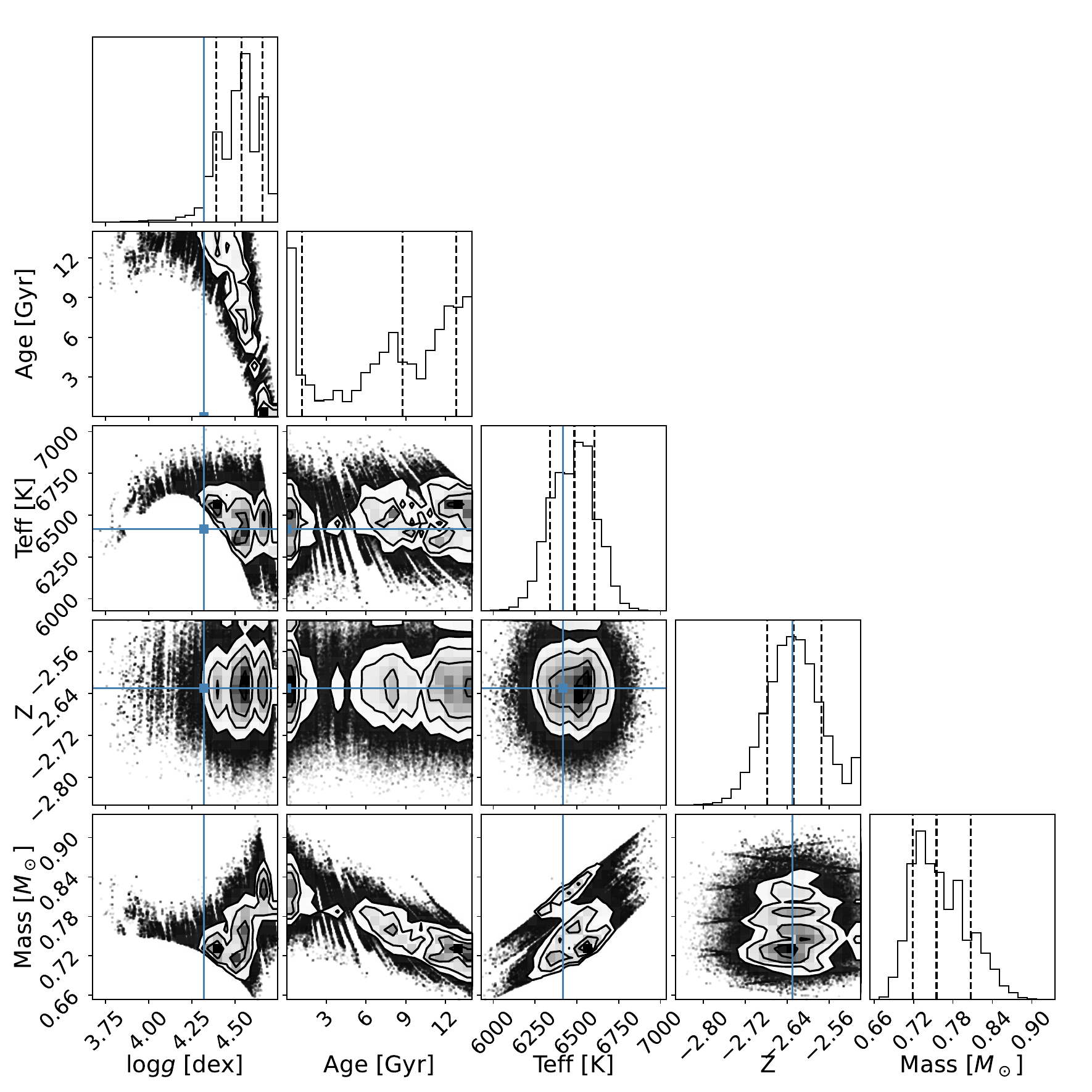}
\includegraphics[width=0.45\textwidth]{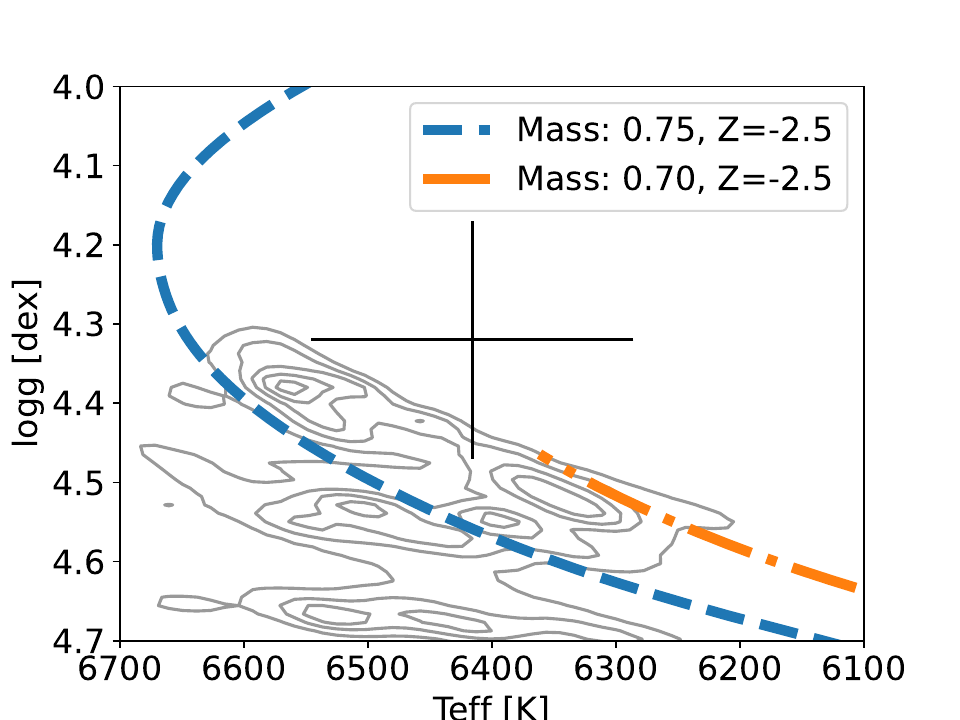}
\caption{Corner plots and evolutionary tracks for stellar parameters of \gstar. \textbf{upper panel}: Posterior distribution of the interpolated values for log $g$, Age, \teff, metallicity, and mass. The blue lines in the corner plot show the observed values for the star.  \textbf{lower panel}: Tracks are shown for comparison for $0.75$ \msun\, and $0.70$ \msun\, at a metallicity of $-2.5$. Grey contours show the sampled distribution.  }
\label{fig:interp_gaia}
\end{figure}

The MIST tracks are shown in Fig.\ref{fig:interp_gaia} Fig. \ref{fig:interp_scholz}.

\section{SED}
\label{C.sed}

\begin{figure}[ht!]
\centering
\includegraphics[width=0.4\textwidth]{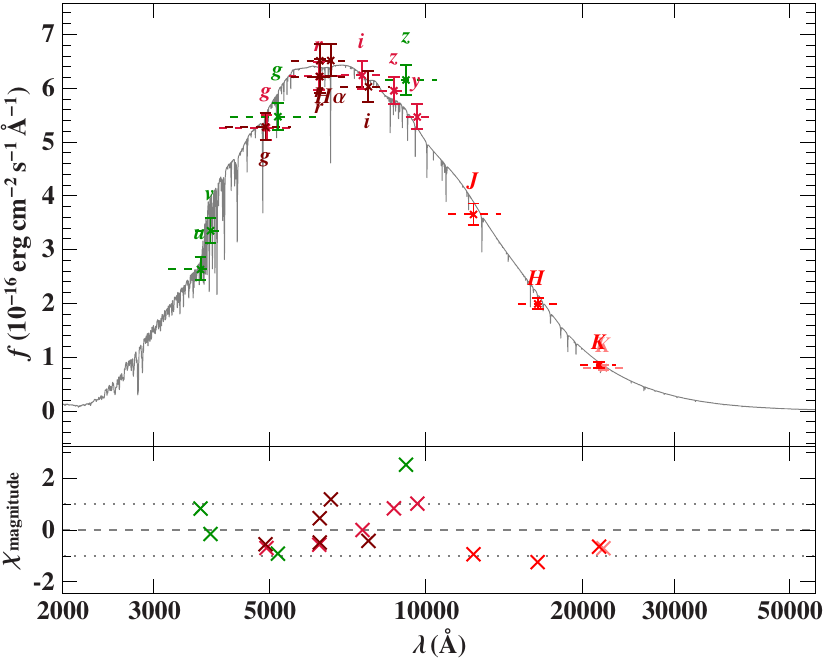}
\caption{Spectral energy distribution of \gstar. The best-fit model is plotted in grey. Photometric values are taken from 2MASS \citep{2mass}, IGAPS \citep{int}, Pan-STARRS \citep{panstarss}, Skymapper DR4 \citep{skymapper}, and UKIDSS DR6 \citep{ukidss}. Residuals are shown below.}
\label{fig:gaiaSED}
\end{figure}

\begin{figure}[h!]
\centering
\includegraphics[width=0.4\textwidth]{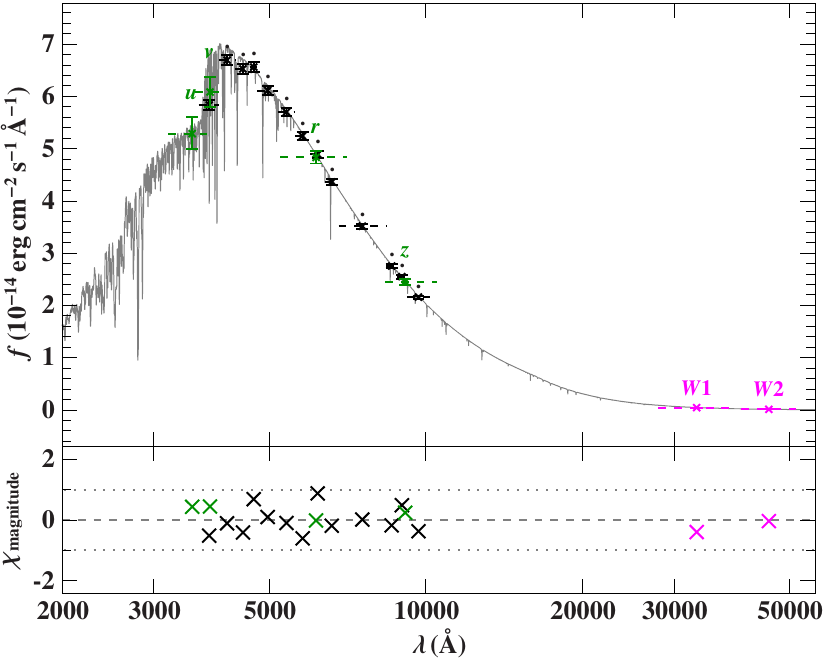}
\caption{The best-fit SED model for \wise\ is shown in grey. Points from Gaia XP spectra \citep{2022yCat.1355....0G}, Skymapper \citep{skymapper}, and Wise \citep{wise} are marked in black, green, and pink. Residuals are shown below. }
\label{fig:scholzSED}
\end{figure}

\begin{figure}
\centering
\includegraphics[width=0.4\textwidth]{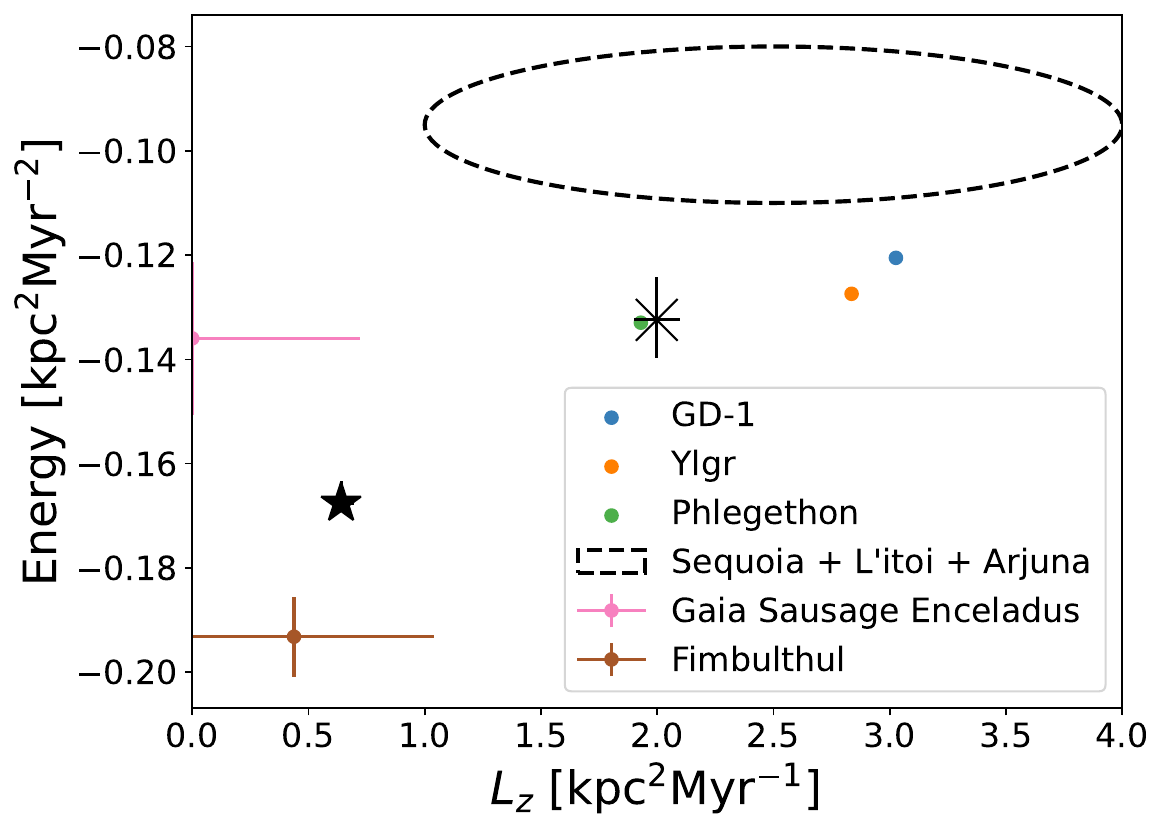}
\caption{Lz-Energy plot of retrograde streams with \feh $<-1.5$, as well as the two stars; \gstar\, marked with the cross, and \wise\, marked with the star. 
}
\label{fig:lz_E}
\end{figure}

\section{Light Curve Modeling}
\label{A.lcmodel}

We modeled the light curve by setting 'irrad\_frac\_refl\_bol' of the primary to 1.0 and using the Horvat scheme \citep{2019ApJS..240...36H}. While the light curve was modelled setting pblum to dataset-scaled, this was later changed by sampling also pblum in the emcee run. The mass of the primary was kept fixed, allowing a better estimate of q. However, the light curve is not precise enough to get a reliable mass fraction, and further RV follow-up is needed to properly solve the system. 
\begin{figure*}
\centering
\includegraphics[width=0.9\textwidth]{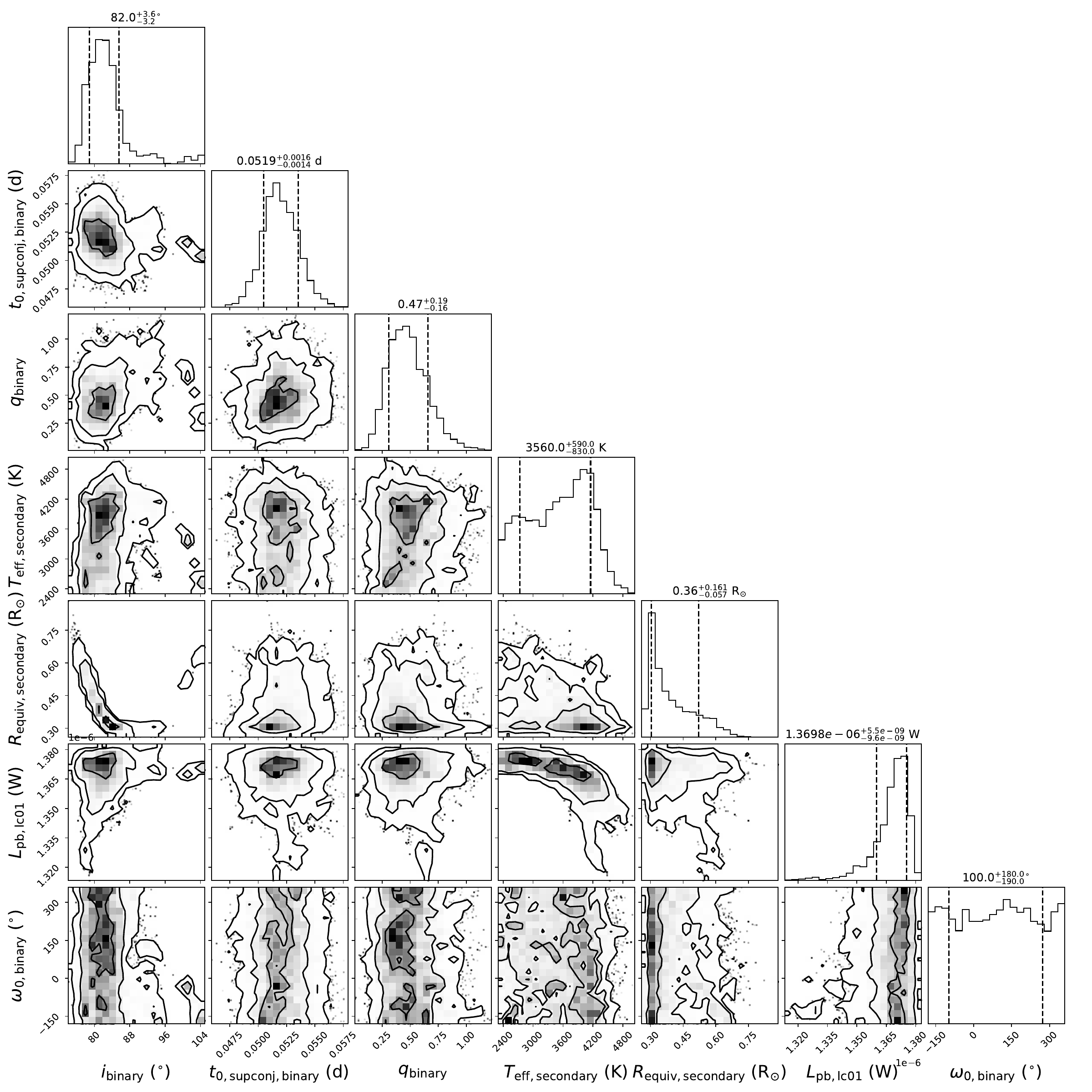}

\caption{Corner plot for the sampled parameters modeled in phoebe.  }
\label{fig:lccorner}
\end{figure*}
\section{Galactic origin of the stars}\label{sect:E.origin}

For \gstar\ the trajectory (see Fig. \ref{fig:kin_bound}) and the $L_z$-energy diagram (Fig. \ref{fig:lz_E}) suggest that the star was accreted in a merger event. Similarly, unlike in-situ halo stars that are predicted to have lower $r_{\rm max}$ \citep{2012A&A...538A..21S}, this star has a much higher $r_\mathrm{max}$. However, the accretion-based origin does not line up well with the low \feh, high \alp\ values of the star. Based on its surface composition, the star resembles the \textit{in-situ}/inner halo, since accreted stars are predicted to have lower $\alpha$ enhancement \citep{2024A&A...682A.116N}. The position of the star in Fig. \ref{fig:lz_E} along with its metallicity shows that it is close to both the region of the Phlegethon stream and nearby the Sequoia merger \citep[values adapted from][]{2021ApJ...909L..26B,2022MNRAS.516.5331M}.

The $L_z$-energy coordinates (see Fig. \ref{fig:lz_E}) place \wise\ between the Gaia-Sausage and close to the Fimbulthul stream, which is the stream coming from the $\omega$ Cen globular cluster \citep[values adapted from][]{2021ApJ...909L..26B,2022MNRAS.516.5331M}. The high \alp\, value therefore implies that it is likely to be of \textit{in-situ} origin and the $L_z$-energy values are merely coincidental.

\end{document}